\documentclass[twocolumn,prb,aps,showpacs,superscriptaddress]{revtex4-1}
\usepackage{graphicx} \usepackage{amsmath} \usepackage{color} \usepackage{soul}
\newcommand{\1}{\textcolor{red}} 
\usepackage[dvipdfm]{hyperref}

\newcommand{\mycol}{1}

\begin{document}

\title {Rabi oscillations of pinned solitons in spin chains: A route to quantum computation and communication.}

\author{S. Bertaina} \email{sylvain.bertaina@im2np.fr}
\affiliation{Aix-Marseille Universit\'{e}, CNRS, IM2NP UMR7334, 13397 cedex 20,
Marseille, France.}

\author{C-E. Dutoit} \affiliation{Aix-Marseille Universit\'{e}, CNRS, IM2NP
UMR7334, 13397 cedex 20, Marseille, France.}

\author{J. Van Tol} \affiliation{Physics Department and the National High
Magnetic Field Laboratory, Florida State University, 1800 E. Paul Dirac Drive,
Tallahassee, Florida 32310}

\author{M. Dressel} \affiliation{Physikalisches Institut, Universit\"{a}t
Stuttgart, 70550 Stuttgart, Germany.}

\author{B. Barbara} \affiliation{Univ. Grenoble Alpes and CNRS, Inst. N\'{E}EL, F-38042
Grenoble, France.}

\author{A. Stepanov} \affiliation{Aix-Marseille Universit\'{e}, CNRS, IM2NP
UMR7334, 13397 cedex 20, Marseille, France.}

%\date{Submitted}

\begin{abstract} We provide the first evidence for coherence and Rabi
oscillations of  spin-solitons pinned by the local breaking of translational
symmetry in isotropic Heisenberg chains (simple antiferromagnetic-N\'{e}el or
spin-Peierls).We show that these correlated spin systems made of hundreds of
coupled spin  bear an overall spin S=1/2 and can be manipulated as a single
spin. This is clearly contrary to all known spin-qubits which are paramagnetic
centres, highly diluted to prevent decoherence. These results offer an
alternative approach for spin-qubits paving the way for the implementation of a
new type of quantum computer. \end{abstract}

\pacs{71.27.+a,03.67.-a,75.10.Pq,76.30.-v}

\maketitle

Most physical, chemical or biological systems showing quantum oscillations are
of relatively small size: isolated NV-centers in diamond \cite{Childress2006a},
4$f$ or 3$d$ transition-metal ions (single-spins, 0.1 nm)
\cite{Bertaina2007,Bertaina2009}, single molecule magnets \cite{Bertaina2008}
(15 spins, 1 nm), or marine algae (5 nm wide proteins) \cite{Collini2010}. Their
environmental couplings are necessarily weak in order to reduce damping
\cite{Leggett1987}. In magnetic systems, decoherence is usually dominated by
spin-bath dipole-dipole interactions \cite{Prokof'ev2000} and observations of
quantum oscillations require {\it qubits dilution}. Here, we report the first
experimental realization of what we call "soliton qubits". Contrary to all
existing qubits, each qubit is made of hundreds of strongly exchange-coupled
spins sitting at defects of strongly correlated spin-chains. This takes
advantage of one of the most remarkable properties of one- dimensional spin
systems: their quite unconventional response to translational symmetry breakings
which often consists in the formation of magnetic defects described as spin
solitons (kinks or domain walls). In as grown single-crystals, these defects are
generally associated with local inhomogeneities such as crystallographic
defects/disorder, bond alternations, chain ends etc. Theoretical prediction for
Heisenberg spin 1/2 quantum spin chains show that each of these magnetic defects
carry an overall spin S=1/2 which can serve as a qubit. The idea of pinned-
soliton  was successfully applied to explain the magnetic
susceptibility\cite{Sirker2007}, NMR\cite{Alet2000} and  some EPR
experiments\cite{Zorko2002} in spin chains, however, to date   almost {\it
nothing is known either experimentally or theoretically about the coherent
dynamics of solitons.}

In this communication, we report for the first time on Rabi oscillations of
pinned spin-solitons in isotropic Heisenberg chains (simple
antiferromagnetic-N\'{e}el or spin-Peierls). We demonstrate that these
collective extended defects  bear an overall spin S=1/2 and can be manipulated
as single quantum spins. This clearly presents an alternative to all known
spin-qubits which are paramagnetic centres in highly diluted samples. These
results offer a new approach for spin-qubits paving the way for the
implementation of a new type of quantum computer.

\begin{figure} \centering \includegraphics[bb=19 20 812
834,width=\mycol\columnwidth,clip]{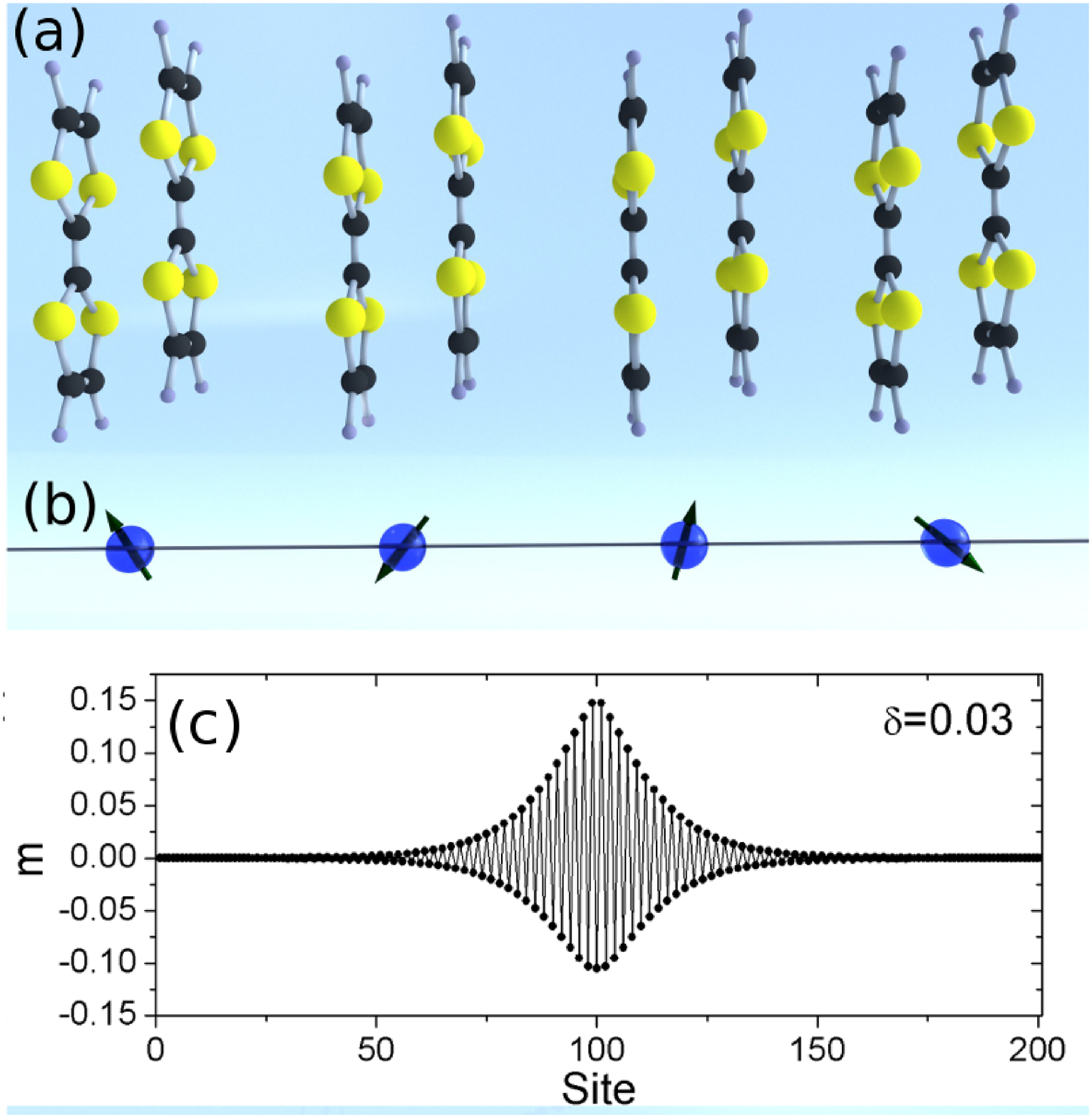} \caption{Colour online.(a)
Schematic representation of (TMTTF)$_2$X chain.  The TMTTF molecules are stacked
in the $a$ axis direction forming the chain. Each double molecules of TMTTF
carries a spin $S=1/2$. The counter anion X is not represented here. (b),
magnetic representation of the chain. The blue spheres are the double molecules
of TMTTF carrying a spin (black arrow) and coupled along the $a$ axis by the
exchange $J$(1$\pm \delta$). (c), site magnetization profile induced by one bond
defect placed in the middle of dimerized chain in the $Mz=1/2$ space computed by
DMRG (see method \cite{Bauer2011}).}\label{fig:1} \end{figure}

The first evidence of Rabi oscillations in sol-qubits is obtained on
single-crystals of the so-called antiferromagnetic quantum spin chains
(TMTTF)$_2$X, with X = AsF$_6$, PF$_6$, SbF$_6$ (Fig. \ref{fig:1}). This family
of organic magnets, also called Fabre salts \cite{Moser1998}, was extensively
studied during the last decades and shows an extremely rich phase diagram
\cite{Jerome1991}. The systems with X = AsF$_6$ and PF$_6$ show a gapped
dimerized spin-pair singlet ground-state below their spin-Peierls transitions at
$T_{SP}=13$ K and 19 K respectively, whereas the system with X = SbF$_6$
exhibits a N\'eel antiferromagnetic phase below $T_N=7$ K.

The Hamiltonian of a S=1/2 Heisenberg chain can be written: \begin{equation}
H=J\sum_i[(1-\delta)S_{2i-1}S_{2i}+(1+\delta)S_{2i}S_{2i+1}] \end{equation}

where $S_i$ are the S=1/2 spin operators, $J$ the exchange coupling and $\delta$
the dimerization parameter. If $\delta$ = 0, (1) describes the uniform
Heisenberg AF chain the ground state is a gapless S=1/2 doublet. This is the
case of the spin-chain system (TMTTF)$_2$SbF$_6$  at T$>$T$_N$. If $\delta > 0$,
(1) describes a Spin-Peierls chain: the ground state is a gapped dimerized
spin-pair singlet (S=0) at temperatures below T$_{SP}$.  This is the case of the
systems (TMTTF)$_2$PF$_6$ and (TMTTF)$_2$AsF$_6$ (if $T>T_{SP}$, $\delta=0$ in
these systems too). The isotropic part of the exchange interaction of these
three systems, $J\sim$ 400 K, is relatively large whereas their intra- and
inter-dimer contributions of (TMTTF)$_2$PF$_6$ and (TMTTF)$_2$AsF$_6$, give a
bond alternation (dimerization) parameter $\delta\sim 0.03$ leading to the
singlet-triplet gap $\Delta =$ 35 K \cite{Dumm2004}. Such a value is more than
enough to provide an excellent separation of the ground state at the Kelvin
scale of temperatures, i.e. an extended collective singlet ground-state in which
two trapped soliton qubits can be strongly entangled.

The single crystals of (TMTTF)$_2$X were grown by an electrochemical
technique\cite{Yasin2012}. The crystals are needle-shaped with typical
dimensions: 3x0.5x0.1 mm$^3$. They crystallize in the triclinic P$\overline{1}$
space group. The magnetic principal axes ($b^\prime$ and $c^*$) are different
from the crystallographic axes and correspond to the extrema of g-factor in the
plan perpendicular to $a$ axis. The static magnetic field can be applied in any
direction in the $b^\prime$$-$$c^*$ plane.   For each set of measurements a
fresh sample was used.

Continuous Wave (CW) and Pulsed Electron Paramagnetic Resonance experiments were
performed with the three systems using a conventional X-band Bruker spectrometer
operating at about 9.6 GHz between 3 K and 300 K and enabling sample rotations.
The crystals were glued on the sample holder with their $a$-axis oriented along
the microwave field ($h_{mw}$) direction which is the same as the
sample-rotation direction (the static $H$ being applied in the basal
$b^\prime$$-$$c^*$ plane with $\theta$ the angle between $H$ and $c^*$ ).

Above 30 K a single Lorentzian-shaped EPR line (main line) is observed,
displaying an anisotropy of $g$ factor, associated with different orientations
of $H$. This anisotropy and the temperature dependence are typical of uniform
quantum Heisenberg spin chains and was intensively studied in the past
\cite{Salameh2011} Below about 30 K a second EPR line, a very sharp one, appears
in the three systems at the same magnetic field as the main line. The integrated
intensity of this sharp signal (SS) is much smaller (by a factor of 10$^2-10^3$)
than the one of the main line, indicating its defect origin.

\begin{figure} \centering \includegraphics[bb=0 0 471
476,width=\mycol\columnwidth,clip]{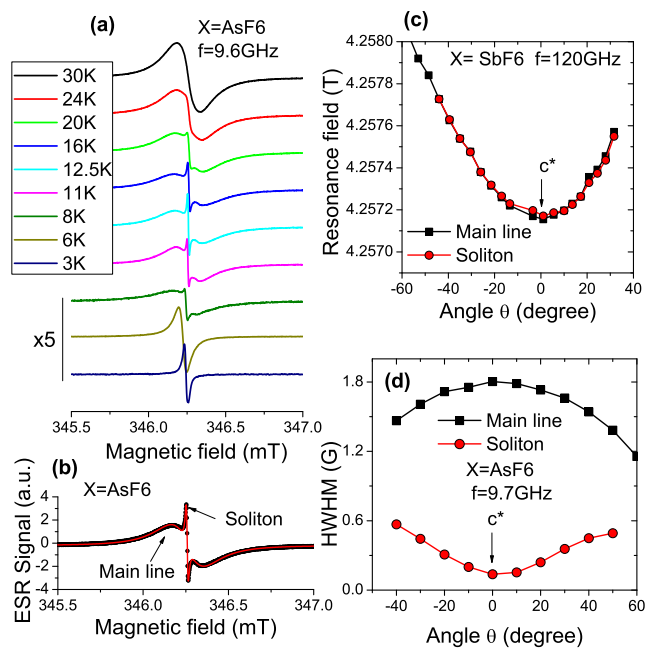} \caption{color online. CW-EPR,
evidence of the soliton signal. (a) Set of CW-EPR spectra of (TMTTF)$_2$AsF$_6$
when the temperature decreases for $H\parallel c^*$. (b) EPR spectrum of
(TMTTF)$_2$AsF$_6$ at $T=12.5$ K. The dots are the experiments and the red line
is a fit of 2 derivative of Lorentzian. (c) Angular dependence of the resonance
fields of (TMTTF)$_2$SbF$_6$ at 120 GHz. The black squares are the main lines,
the red circles are the soliton lines. (d) Angular dependence of the linewidth
of (TMTTF)$_2$AsF$_6$ at 9.5 GHz and $T=12.5$ K.} \label{fig:2} \end{figure}

As an example, figure \ref{fig:2}(a) gives a set of spectra obtained with X =
AsF$_6$ between 30 K and 3 K with $H\parallel c^*$ showing how the very sharp
signal progressively appears and becomes dominant when the temperature
decreases. Below $T_{SP}=13$ K the system starts to dimerise and enters in the
spin-Peierls phase. The intensity of the main peak drops but that of the SS
remains almost unchanged. The same behaviour has been observed in PF$_6$
compound. In the SbF$_6$ system, the linewidth of the main line diverges when
$T$ decreases down to $T_N=7$ K whereas the SS remains almost unchanged, except
below $T_N$ where it disappears \cite{Coulon2004}. This behaviour is contrary to
that of an isolated paramagnetic impurity, the intensity of which is strongly
temperature-dependent (Curie law), and is rather characteristic of an ensemble
of correlated spins.

In the limit of the resolution of the X-band spectrometer, the measured
resonance field and therefore the $g$ factor were identical for both the broad
and sharp peaks and did not change significantly for different molecules. A
two-lorentzian fit shows that their linewidths differ by a factor of ten ($\sim$
1 G for the former and $\sim$ 0.1 G for the latter, Fig. \ref{fig:2}(b)). In
order to improve the resolution we measured the CW-EPR in these systems at 120
GHz in a home-made quasi-optical CW-EPR spectrometer \cite{VanTol2005} (the
angular dependence of the magnetic field resonance of the SbF$_6$ system is
given in figure \ref{fig:2}(c). The resonance fields of both lines remain
identical for all the applied-field orientations, to an accuracy better than
10$^{-5}$.

Figure \ref{fig:2}(d) gives the linewidth angular dependence of the two peaks
observed. The broad/sharp linewidth ratio reaches its maximum of $\sim$10 when
the static field is applied along the $c^*$ axis. Surprisingly, whereas the
resonant field of the two peaks is observed to be the same whatever the angle
$\theta$, the angular dependences of their linewidths are opposite.  The width
of the broad line follows the well-known behavior of a uniform $S=1/2$ chain,
proportional to $1+\cos^2\theta$, with a maximum for $\theta=0^{\circ}$, whereas
the one of the sharp line is minimum for $\theta=0^{\circ}$ as expected for a
line issued from defect-induced correlated spins.

To sum up, while the SS has quite unusual,  compared to a isolated impurity
signal, the temperature and the linewidth angular dependencies, one of the most
important features which supports its spin soliton origin is the finding that SS
g-factor is {\it precisely the same as the one of the main EPR line}. This
result is perfectly consistent with a physical picture which one can suggest to
describe the defect formation in TMTTF charge transfer salts. As we have
mentioned above any translation symmetry breaking of molecular structure in the
direction of the chain, crystallographic or electron charge inhomogenities, will
lead locally to intra-chain exchange modification which in turn will cause a
spin soliton formation close to this defect. The observed EPR signal in this
case will come not from an impurity centre itself (a localized electron
somewhere in TMTTF molecule) but from a collective precession of hundreds of non
perturbed spins of the chain which form a soliton. Quite naturally this
precession is characterized by the same g-factor as the main EPR signal.

Before going further in our experimental investigation, let us explain in more
details the condition of appearance and stability of solitons near bond defects
in spin chains. For that we have  performed numerical DMRG studies of
alternating spin chains in which the successive bonds are J1 =J(1+$\delta$) , J2
=J(1-$\delta$) with J1$>$ J2. In fig. \ref{fig:1}c a typical distribution of the
soliton magnetic moment is shown for a defect characterized  by the succession
of two strong bonds placed in the middle of the chain. The result, shown in Fig
\ref{fig:1}, fully corresponds to a spin-soliton, and after integration we find
that it carries a $S=1/2$ These results are in full agreement with similar
studies previously published\cite{Nishino2000a} and \cite{Nishino2000} for
comprehensive theoretical studies.

\begin{figure} \centering \includegraphics[bb=10 0 480
480,width=\mycol\columnwidth,clip]{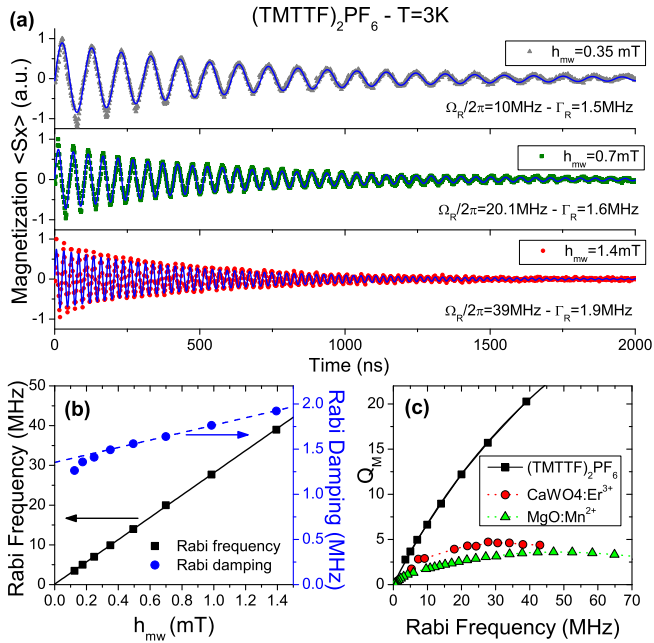} \caption{colour online. Rabi
oscillations and coherence properties of soliton-qubit. (a) Series of Rabi
oscillations of (TMTTF)$_2$PF$_6$ measured at $T=3$ K and $f=$9.7 GHz using the
FID method while increasing the microwave magnetic field $h_{mw}$. Each point is
an average of 1000 FID measurements. The blue lines are fits using
$\sin(\Omega_R).\exp(t/\tau_R)$. (b) Rabi frequencies ($\Omega_R/2\pi$) and
damping ($1/\tau_R$) as function of $h_{mw}$. The black line represents the Rabi
frequency of spin S=1/2.  The blue line is a fit using $\Gamma_0+\gamma.h_{mw}$,
$\Gamma_0=1.35$ MHz (the zeros microwave field coherence) and $\gamma=0.41$
MHz/mT the effect of microwave field on the Rabi damping. (c) Merit factor $Q_M$
of (TMTTF)$_2$PF$_6$ compared to diluted ion systems \cite{DeRaedt2012}. The
black line is a simulation using the $\Gamma_0$ and $\gamma$ from (b). }
\label{fig:3} \end{figure}

The pulsed EPR experiments were performed on (TMTTF)$_2$AsF$_6$ and
(TMTTF)$_2$PF$_6$ single crystals with a microwave field $h_{mw}$ varying
between 0.1 and 1.5 mT. The coherent signal, resulting from the SS observed in
CW experiments, was recorded by the Free Induction Decay method. %\cite{livre}.
 The spin-echo detection cannot be used here, in contrast with most other known
 systems, because of the absence of sizable inhomogeneous line-broadening. In
 fact the line-broadening of $\sim$ 0.1 G observed is essentially homogeneous
 (see discussion below). Examples of Rabi oscillations obtained at 3 K for X =
 PF$_6$ are given in figure \ref{fig:3}. The oscillations are very well-fitted
 by the exponentially damped sinusoidal function $\left\langle
 S_x(t)\right\rangle \propto \sin(\Omega_R t)\exp(-t/\tau_R)$ with $\Omega_R$
 the Rabi pulsation and $\tau_R$ the Rabi damping characteristic time. The Rabi
 frequency increases linearly with $h_{mw}$ with a slope
 $d(\Omega_R/2\pi)/dh_{mw}\sim$ 28 MHz/mT close to the expected value for spins
 $S=1/2$ (Fig. 3(b)), thus providing an additional  evidence in favour of our
 model of a  trapped soliton with $S=1/2$. This figure also shows the
 microwave-field dependence of the Rabi damping 1$/\tau_R$ allowing one to
 evaluate the damping by the microwaves
 \cite{Bertaina2007,Bertaina2008,Shim2012}. This ``over-damping", associated
 with inhomogeneous line-broadening due to the distributions of the transverse
 $g$ factor (resulting itself from weak ligand-field distributions) or to the
 microwave-field amplitude (non-homogeneous cavities) is generally unavoidable
 \cite{DeRaedt2012}. In the present case, and this is rather exceptional, the
 ``over-damping'' is particularly small due to the homogeneous character of the
 EPR line. The measured value of $d(1/\tau_R)/dh_{mw}$ = 0.4 MHz/mT (Fig. 3(b)),
 is 10 to 50 times smaller than in ion-diluted systems
 \cite{Shim2012,DeRaedt2012}.  In addition, contrary to most other systems, the
 figure of merit $Q_M= \Omega_R\tau_R/2\pi$ of sol-qubits in (TMTTF)$_2$PF$_6$
 does not saturate when the Rabi frequency increases (Fig. 3(c)) and follows the
 expression $Q_M = \Omega_R/(8.5+0.015 \Omega_R)$. Our largest microwave field
 $h_{mw} = 1.5$ mT gives at 3 K $Q_M \sim 23$ while a value of the order of 70
 is expected for a larger field. Whereas $Q_M$ of CaWO$_4$:Er$^{3+}$ and
 MgO:Mn$^{2+}$ which saturate at about 3, $Q_M$ of sol-qubits is larger by an
 order of magnitude.

The fact that sol-qubits are almost not sensitive to system parameters ($g$
factors, inhomogeneous magnetic and microwave fields, etc.) is worthy of special
consideration. Indeed, with all the other types of qubits these distributions
lead to inhomogeneous distributions of Rabi frequencies (the qubits are not
identical) giving destructive interferences and decoherence. This is a major
roadblock for the implementation of a spin-based quantum computer. Magnetic
dipole-dipole interactions are also an inevitable and constitute a source of
decoherence unless the qubits are very far from each others preventing any kind
of manipulation. With  sol-qubits, the situation is just opposite: the strong
spin exchange interactions (J=400 K) eliminates decoherence through the well
known exchange narrowing mechanism, as this is shown in the S=1 Haldane spin
chains \cite{Mitra1992} picture developed for interacting S=1/2 degrees of
freedom. This explains why the coherence of sol-qubits is robust against
microwaves even at high power. Finally these long-living sol-qubits, even if
they are distant, are easily coupled to each others and controlled through
effective isotropic exchange interaction along the spin-chain.

In conclusion, by observing long-living Rabi oscillations of sol-qubits in
Heisenberg gapped spin-Peierls systems, we provide first evidence for coherence
in spin chains and more particularly in solitons trapped at exchange defects in
spin-chains. Due to isotropic inter-qubit exchange interaction, the EPR lines
observed are homogeneous and narrowed, eliminating all the usual decoherence
mechanisms such as the one associated with non-perfectly identical qubits and
dipolar interactions.

Following the idea of spin-qubit quantum computer\cite{Loss1998}, an
increasing number of proposals were made during the last decades showing
theoretically how spin chains may enable the implementation of a quantum
computer by  using them as quantum wires to connect distant qubit registers
without resorting to optics
\cite{Wootters1998,Arnesen2001,LagmagoKamta2002,Bose2003,CamposVenuti2006,Amico2008}.We speculate that our sol-qubits might be ideal candidates for the
realization of such a computer since they represent intrinsic spin registers
which do not require the addition of any supplementary spin to the system and
since they perfectly match the communication channel - spin chain.

We acknowledge the city of Marseille, Aix Marseille 
Universit´e, NHMFL user program, and IR RENARD FR3443 
for financial support. The NHMFL is supported by NSF Cooperative Agreement No. DMR-1157490, the U.S. Department of Energy, and by the State of Florida. We thank G. Gerbaud and PFM St. Charles for technical support. S.B. and B.B. thank 265 I. Chiorescu and S. Miyashita for valuable discussions.

\end{document}